\documentclass[preprint,3p,times,twocolumn]{elsarticle}

\usepackage{graphicx}

\journal{Physics Letters B}

\begin{document}

\begin{frontmatter}

\title{Observation of near-scission ``polar" and ``equatorial"  proton emission in heavy-ion induced fission}
\author[BARC,HBNI]{Pawan Singh}
\author[BARC,HBNI]{Y. K. ~Gupta\corref{cor1}}
\cortext[cor1]{Corresponding author, Email:ykgupta@barc.gov.in}

\author[BARC]{G. K. Prajapati}
\author[BARC]{B. N. Joshi}
\author[SVNIT]{V. G. Prajapati}
\author[BARC]{N. Sirswal}
\author[BARC]{K. Ramachandran}
\author[HYD]{A. S. Pradeep}
\author[HPD]{V. S. Dagre}
\author[IUAC]{M. Kumar}
\author[IUAC]{A. Jhingan}
\author[PPSU]{N. Deshmukh}
\author[BARC]{B. V. John}
\author[BARC,HBNI]{B. K. Nayak}
\author[BARC]{D. C. Biswas}
\author[BARC]{R. K. Choudhury}
\address[BARC]{Nuclear Physics Division, Bhabha Atomic Research Centre, Mumbai - 400085, INDIA}
\address[HBNI]{Homi Bhabha National Institute, Anushaktinagar, Mumbai 400094, India}
\address[SVNIT]{Department of physics, Sardar Vallabhbhai National Institute of Technology, Surat -395007, India}
\address[HYD]{Department of Physics, Hyderabad Institute of Technology and Management, Hyderabad, 501401, India}
\address[HPD]{Health Physics Division, Bhabha Atomic Research Centre, Mumbai 400085, India}
\address[IUAC]{Inter-University Accelerator Centre, Aruna Asaf Ali Marg, New Delhi 110067, India}
\address[PPSU]{School of Sciences, P P Savani University, Dhamdod, Kosamba, Surat - 394125, Gujarat, India}

\date{\today}

\begin{abstract}
Proton and $\alpha$-particle energy spectra were measured in coincidence with fission fragments at different relative angles in  $^{16}$O (96 MeV)  + $^{232}$Th reaction. The multiplicity spectra were analyzed within the framework of a Moving Source Disentangling Analysis (MSDA) to determine contributions from different emission stages. The MSDA conclusively shows ``Near Scission Emission (NSE)" as an essential component in the multiplicity spectra. In contrast to NSE $\alpha$ particles which emit mainly perpendicular (``equatorial emission"), the NSE protons are observed to be emitted perpendicular as well as parallel (``polar emission") to the fission axis with similar intensities ($\sim$20\% for each). Thus, around 40\% of total pre-scission protons are emitted near the scission stage, whereas the same fraction for  $\alpha$ particles is only around 10\%. The inevitable presence of ``polar" and ``equatorial" NSE protons in a heavy-ion induced fission has been observed for the first time. Present results opens up a new avenue to study the heavy-ion induced fission dynamics.



%



\end{abstract}

\end{frontmatter}

Understanding about the precise nature of the nuclear viscosity (one- versus two-body) and its dependence on temperature and coordinate space (deformation) is still quite unclear \cite{Blocki1978, Davies1976}. During the nuclear fission,  the nuclear shape evolves from equilibrated mono-nucleus to the highly deformed scission stage, involving a large scale mass flow in a dissipative medium. It provides an opportunity to learn about the properties of nuclear viscosity \cite{Hilscher_Rossner_1992}. During fission, the finite nuclear matter undergoes through a steep potential gradient. As debated earlier, the energy dissipation at the scission stage might be quite different than just before it \cite{Simenel2014, Rizea2013}. The rapidly moving potential walls  might justify suitability of one- over two-body viscosity near to the scission point \cite{Blocki1978, Davies1976}. A clear understanding about the nuclear viscosity at such a nascent stage is of fundamental importance, in particular with varying temperature. The scission time delay, i.e., the time from neck rupture point at the finite radius to the absorption of the neck stubs by the fragments, connects directly with the nuclear viscosity at such a subtle stage.

The neck rupture process has been debated quite actively in the past \cite{ZXRen2022PRL, JBNatowitz2023}. Particle emission near the scission stage, termed as Near Scission Emission (NSE) provides finer details about the underlying mechanism of the neck rupture process. Low energy fission (spontaneous-, thermal-neutron induced, photo-fission) can be employed for studies related to the neck rupture process at low temperatures. In the low energy fission, it is widely accepted that the neck rupture is quite sudden \cite{nadkarni, halpern}. Yield of various Light Charged Particles (LCPs) has been measured in the low energy fission \cite{nadkarni, halpern}. It is seen that among various LCPs, $\alpha$-particle emission near the scission stage is the dominant one. Historically, in the low-energy fission these  $\alpha$ particles  are also known as Long Range $\alpha$ particles (LRAs) \cite{Rutherford1916}. Very recently, studies related to these LRAs have been extended in neutron-deficit heavy mass region and superheavy nuclei \cite{Khuyagbaatar2024}.

The NSE $\alpha$ particles are preferentially emitted perpendicular to the fission axis, also known as ``Equatorial Emission (EE)" \cite{Polar1980}. A very small fraction is also emitted along the fission axis, referred as the ``Polar Emission (PE)" \cite{Polar1980, PIASECKI1982, PIASECKI1970, PIASECKI1975}. The total yield of each $Z$=1 particle is much lesser than $\alpha$ particles, however, their relative intensities are greater in PE than EE \cite{Polar1980, PIASECKI1982}. The difference is striking for the case of protons, where PE component is observed to be around twenty times of that of EE component \cite{PIASECKI1982}. Except $\alpha$ particles, none of the other LCPs has been observed so far to be emitted near the scission stage in heavy-ion induced fission.

Disentangling of NSE particles in heavy-ion induced fission is quite challenging task due to presence of emissions from different stages. In heavy-ion induced fusion-fission, particle emission takes place continuously; from the onset of the fusion process to the stage where produced fission fragments have attained their asymptotic velocities. In addition to the near scission emission, the rest could be categorized into two major groups, namely emissions from the fully equilibrated compound system (pre-scission) and from the fully accelerated fission fragments (post-scission) \cite{Hilscher_Rossner_1992, ykg11, Hinde1989, Lestone1993, Nitto2018}.

$\alpha$-particle emission near the scission stage has been studied quite systematically in a wide energy regime from low-energy to heavy-ion induced fission \cite{nadkarni, halpern, ykg11, Lestone1993, KR2006,  sowonski, ykg22,  schad984, siwek, lindl1987}. It has been conjectured from NSE $\alpha$ particles that the neck rupture becomes slower in going from low energy fission to heavy-ion induced fission (high excitation energy); nuclear viscosity at the scission stage undergoes a changeover from very low to a high value in going from low to high temperatures \cite{ykg11}. It is of fundamental importance to investigate the validity of aforesaid conjecture using different particle emissions from the scission-stage.

Mutual Coulomb forces of the nascent fission fragments close to the scission configuration make the neck region to be very neutron rich.  Absence of Coulomb barrier suggests that neutron emission from the neck should be dominant among all the particles. Theoretical descriptions through the time-dependent Schr\"{o}dinger equation with time-dependent neutron-nucleus potential shows that a large fraction of neutron multiplicity comes from the Scission Neutrons (SN) \cite{Carjan2013, Carjan2015, Carjan2019, Wada2015}. It is also predicted that maximum yield of the SN is along the fission axis (polar emission) due to attractive nuclear force of the nascent fission fragments at the scission stage \cite{Wada2015}. However, experimentally it has not been possible so far to disentangle SN from the total neutron yield in heavy-ion induced fission. 

On the other hand, the Coulomb forces due to nascent fission fragments on the potential third charged particle to be emitted from the neck region (EE) make its energy and angular distributions to be significantly different from the emissions at other stages (pre- and post-scission emissions) \cite{nadkarni, halpern}. Nevertheless, unlike to $\alpha$-particle emission, studies for other LCPs have been quite limited, primarily owing to their low multiplicities. The next candidate after $\alpha$-particle which has been studied in heavy-ion induced fusion-fission to some extent is the proton \cite{KR2006, ikezoe1,ikezoe1994}. However, in the case of proton emission the background contribution to the energy spectra is significantly larger than the $\alpha$-particle. The primary source of the background protons stems from direct reactions with hydrocarbon impurities deposited on the target during the experiments. Amidst these difficulties, the rare events of proton emission near the scission-stage have not been observed so far in any of the heavy-ion induced fission reactions. 

Besides the aforementioned importance of the near-scission component, there is a renewed interest in the pre-scission yield of different particles. In our recent work \cite{ykg22}, it is shown that the $\alpha_\mathrm{pre}$ makes a changeover from high to a very low value in going from asymmetric to a relatively symmetric entrance channels \cite{ykg22}. Such a discontinuous behavior has not been observed in pre-scission neutron multiplicity data. In this context also, it is crucially important to develop proton emission as a probe to study the heavy-ion induced fission.

We have measured the light charged particle energy spectra in coincidence with fission fragments in $^{16}$O + $^{232}$Th reaction at a beam energy of 96 MeV by using a state of the art experimental setup. The results on $\alpha$-particle multiplicity spectra have been presented earlier in Ref. \cite{ykg22}, which revealed a new signature of non-equilibrium fission. In this Letter, we report results on the clear observation of near-scission proton emission in the above mentioned reaction. It is shown that ``polar" as well as ``perpendicular" NSE components are essential ingredients in the proton multiplicity spectra.

The experiment was performed  using $^{16}$O (96 MeV) beam  from BARC-TIFR 14-MV Pelletron accelerator facility at Mumbai. A 
self-supporting metallic foil of $^{232}$Th ($\sim$1.6
mg/cm$^{2}$) was used as the target. Fission fragments (FFs) produced in the reaction were detected using two large area Multi-Wire Proportional Counters (MWPCs) \cite{MWPCNIM2018}, placed in folding angle configuration. The MWPCs were placed at $\theta_\mathrm{f}$=60$^{\circ}$ ($\phi$=0$^{\circ}$) and 100$^{\circ}$($\phi$=180$^{\circ}$) with angular openings of 27.9$^{\circ}$ and 42.3$^{\circ}$, respectively. Fission events were clearly separated from other reaction products by plotting cathode pulse height from one MWPC against the other. The folding angle and azimuthal angle distributions for the FFs peak at around 159$^{\circ}$ and 180$^{\circ}$ which are consistent with the two-body kinematics. The fission fragments detected by the forward ( $\theta_\mathrm{f}$=60$^{\circ}$) and backward ($\theta_\mathrm{f}$=100$^{\circ}$) angle MWPCs will be referred as FF1 and FF2, respectively, henceforth.



\begin{figure}
\centering\includegraphics [trim= 0.11mm 0.5mm 0.1mm 0.1mm, angle=360, clip, height=0.16\textheight]{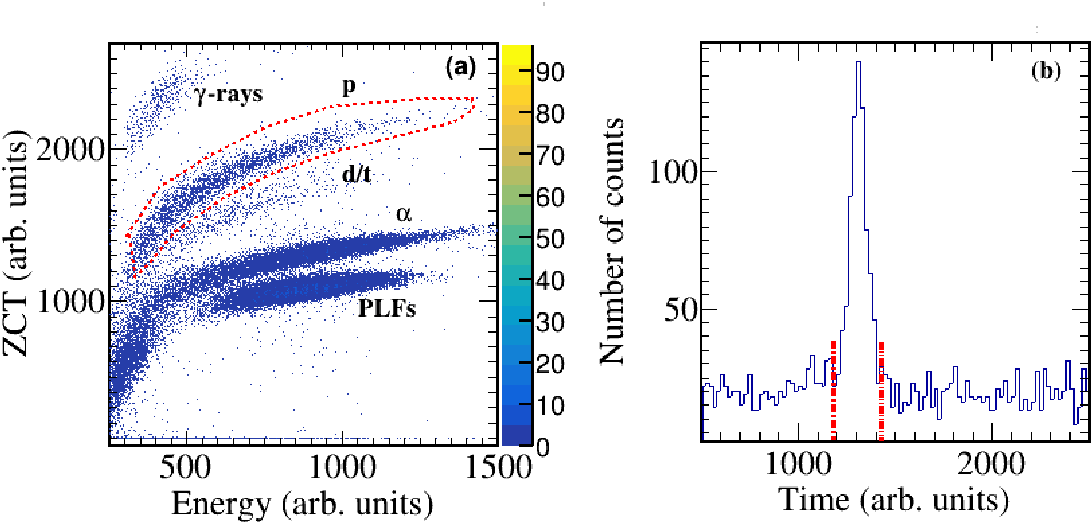}
\caption{
(a) A two dimensional plot of ZCT (Zero Crossover Time) versus energy from a CsI(Tl) detector at a laboratory
angle of 157.8$^\circ$ for different particles produced in the reaction. (b) The time correlations between fission events and proton (see text). The two vertical lines (red-colored) represent the prompt coincidence region.}
\label{PID_ZCT}
\end{figure}

The protons emitted in the reaction were detected by eleven collimated CsI(Tl)-Si(PIN) detectors. Each detector had an angular opening of $\pm$4$^\mathrm{\circ}$. Zero Crossover Time (ZCT) \cite{ykg11, ykg22, ykg12} for the five detectors and Ballistic Deficit (BD) \cite{ykg22, akhil_CsI} for the six detectors were used for the particle identification. A Particle ID (PID) was generated in the case of BD technique \cite{ykg22}. The $\gamma$-rays, light charged particles ($p$, $d/t$, $\alpha$), and the Projectile-Like  Fragments (PLFs) were well separated in the two-dimensional plot of PID (or ZCT) versus energy. A typical particle identification plot (ZCT versus energy) is shown in the Fig.~\ref{PID_ZCT} (a) for a CsI(Tl) detector placed at 157.8$^{\circ}$.

Out of the six CsI(Tl) detectors for which BD technique was used, four were having the same angles as the ZCT detectors with respect to beam as well as the fission fragment direction. An average was obtained of energy spectra from those detectors which have common angles. Thus, effectively a total of seven detectors having unique angle values with respect to the beam direction were considered for the further data analysis. The angle values of these detectors were as $\theta_{p}$=72.1$^{\circ}$ ($\phi$=180$^{\circ}$), 83.9$^{\circ}$ ($\phi$=0$^{\circ}$), 93.0$^{\circ}$ ($\phi$=0$^{\circ}$), 113.9$^{\circ}$ ($\phi$=0$^{\circ}$), 133.8$^{\circ}$ ($\phi$=180$^{\circ}$), 142.8$^{\circ}$ ($\phi$=180$^{\circ}$), and 157.8$^{\circ}$ ($\phi$=0$^{\circ}$). The time correlations between the FFs detected by both the MWPCs were recorded by using a time-to-amplitude converter. Coincidence between the two MWPCs defined the fission-single event. A total of about 10$^{8}$ fission single events were recorded. The time correlations between the fission events and proton were recorded using another time-to-amplitude converter as shown in the Fig.~\ref{PID_ZCT} (b) for a detector placed at 157.8$^{\circ}$, where the two vertical lines (red) represent the prompt coincidence region.

The scintillation light yield response of the CsI(Tl) detectors, measured earlier through in-beam experiments for protons as well as $\alpha$ particles, was employed for proton energy calibration. The light yield response for $\alpha$ particles was measured using $^{12}$C ($^{7}$Li, $\alpha $)  $^{15}$N$^{*}$ and $^{12}$C ($^{12}$C,
$\alpha $)  $^{20}$Ne$^{*}$ reactions  which populate discrete $\alpha$-particle energies. Elastic scattering of protons off the several heavy targets such  as $^{197}$Au, $^{209}$Bi, and $^{232}$Th, were measured to get the light yield response for protons. Also, $^{11}$B beam was bombarded on the Mylar foil with varying beam energies.  The light yield ratio of proton-to-$\alpha$-particle as a function of energy, thus obtained, is observed to be mostly independent of the detector. Nevertheless, an average of the light yield ratio (p/$\alpha$) for three detectors was used for proton energy calibration in the energy range of 5 to 20 MeV. All CsI(Tl) detectors were calibrated periodically throughout the experiment for $\alpha$ particles in the energy range from 4.8 to 8.4 MeV using $^{229}$Th $\alpha$-source. Using earlier measured light yield ratio (p/$\alpha$) and present calibration for $\alpha$ particles provided the calibration for protons.

\begin{figure}
\centering\includegraphics [trim= 0.11mm 0.5mm 0.1mm 0.1mm, angle=360, clip, height=0.19\textheight]{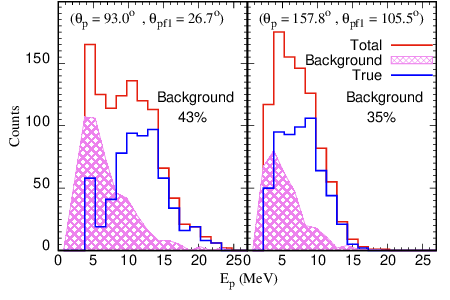}
\caption{
Total (true + background), background, and true proton energy spectra. The total (red solid line) and background (hatched region) spectra correspond to prompt and random coincidences, respectively (see Fig. \ref{PID_ZCT}(b)). The true spectrum (blue solid line) is obtained after subtracting the background spectrum from total spectrum. In the left panel, it is for a relatively forward angle, whereas in the right panel, it is 
for a backward angle (see text).}
\label{BackgroundSub}
\end{figure}

Total (true+background) and only background proton energy spectra, corresponding to prompt and random coincidences (see Fig. \ref{PID_ZCT} (b)), respectively, were obtained. The background contribution is observed to be in the range of 35\% to 45\%. It is more at the forward than at the backward angle CsI(Tl) detectors. A true proton energy spectrum is obtained after subtracting the background from the total spectrum as shown in the Fig. \ref{BackgroundSub}. After correcting for random coincidences, the normalized proton multiplicity spectra were obtained by dividing the coincidence energy spectra  with total number of fission single events.

\begin{figure}[t]
\centering\includegraphics [trim= 0.15mm 0.5mm 0.1mm 0.1mm, angle=360, clip, height=0.55\textheight]{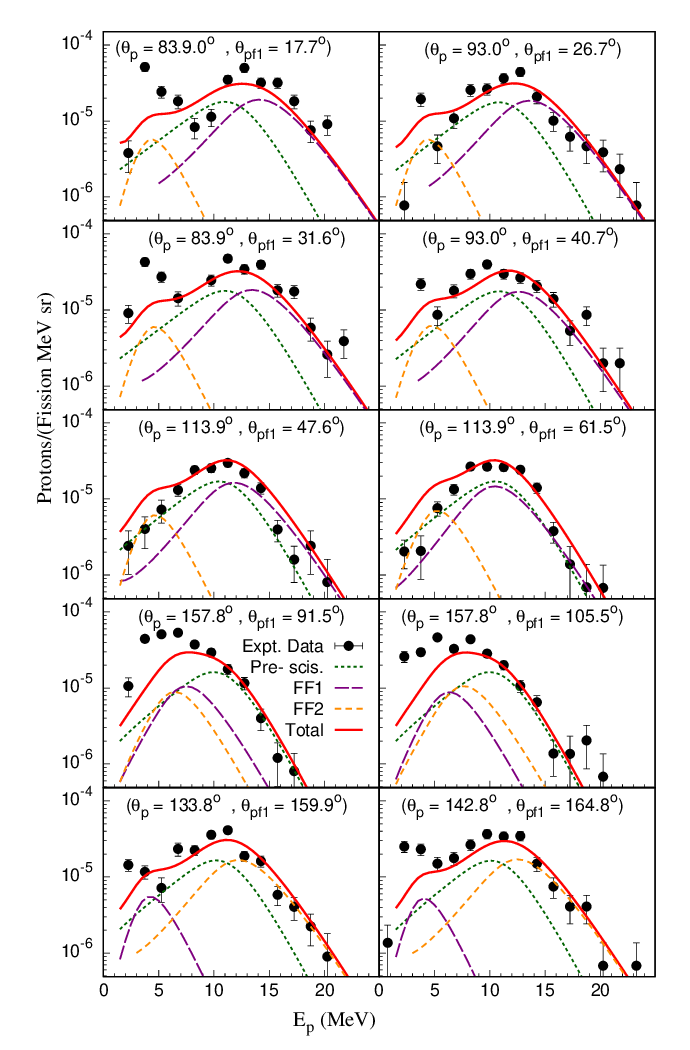} \caption{Proton multiplicity spectra along with fits of moving source model for different combination of laboratory angles of CsI(Tl) detectors with respect to the beam direction, $\theta_\mathrm{p}$ and one of FFs going in the forward direction (FF1), $\theta_\mathrm{pf1}$ in $^{16}$O (96 MeV) + $^{232}$Th reaction. The dotted, long-dashed, and short-dashed curves are contributions from compound nucleus, FF1 (forward angle),  and FF2 (backward angle), respectively. The solid curve in each panel indicates total contribution from all the three sources.}
\label{DataFit}
\end{figure}
\begin{figure}[t]
\centering\includegraphics [trim= 0.15mm 0.5mm 0.1mm 0.1mm,
angle=360, clip, height=0.55\textheight]{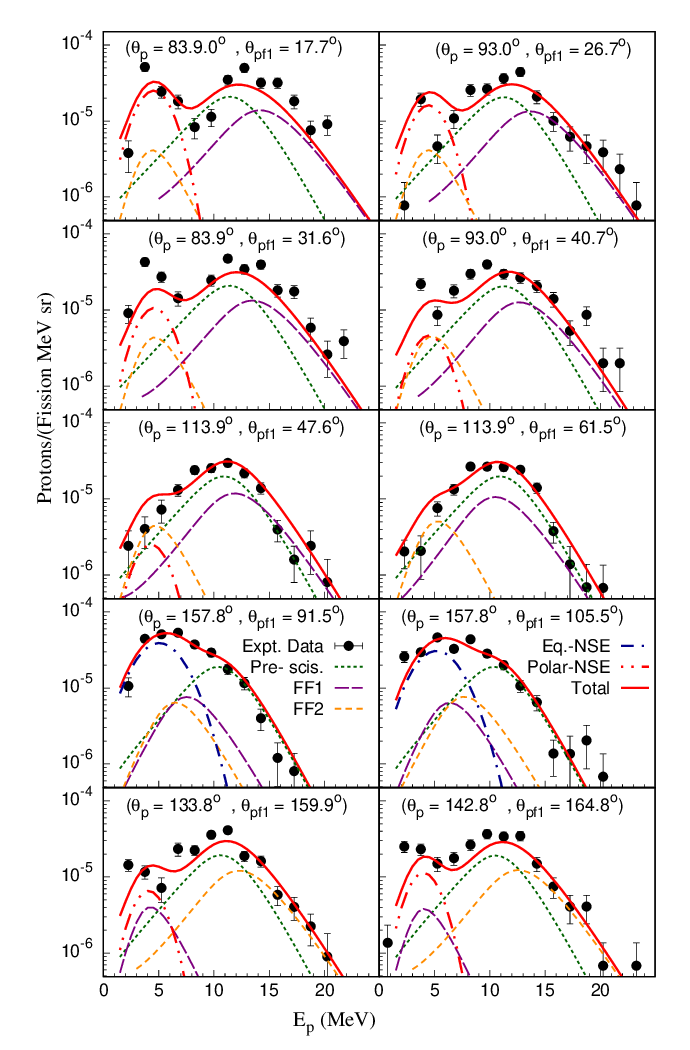} 
\caption{Same as Fig. \ref{DataFit}, but after including the near scission emission (NSE) components, polar (PE) and equatorial (EE). The dotted (green), long-dashed (purple), short-dashed (orange), dash-dotted (blue), and dash-double-dotted (red)  curves are contributions from compound nucleus, FF1 (forward angle), FF2 (backward angle), NSE-EE, and NSE-PE, respectively. The solid curve in each panel indicates total contribution from all the five sources.} 
\label{WithNSE_DataFit}
\end{figure}

The MWPC placed at $\theta_\mathrm{f}$=60$^{\circ}$ was selected as the reference detector for determining relative angle between fission fragment direction and proton in the laboratory frame. The fission events were divided into two angular segments. Thus, for the reference fission detector of $\theta_\mathrm{f}$=60$^{\circ}$ having total angular opening of 27.9$^{\circ}$, each segment was of $\pm$7.0$^\mathrm{\circ}$ angular width. Considering seven widely positioned charged-particle detectors and two angular segments of the fission events, a total of 14 combinations of proton multiplicity spectra, with each having different relative angles with respect to the FF1 ($\theta_\mathrm{p f1}$) and the beam ($\theta_\mathrm{p}$), were obtained. Typical 10 proton multiplicity spectra out of 14 are shown in the Fig. \ref{DataFit}.

A Moving Source Disentangling Analysis (MSDA) is first carried out by including two usual sources; pre- and post-scission emissions. The proton energy spectra in the rest frames for pre- and post
scission sources are calculated using  the constant-temperature level-density formula with the expression \cite{ykg11, KR2006, ykg12};
\begin{equation}
n(\epsilon) =N \pi_{p} \epsilon \sigma(\epsilon) exp(\frac{-\epsilon}{T}) ,
\label{eq:pre-post}
\end{equation}
 where,  $\pi_{p}$ and $ \epsilon$ are the multiplicity and energy of the emitted protons in the rest frame, $T$ is the
 temperature of the source, $\sigma(\epsilon)$ is the inverse reaction cross
section and $N $ is the normalization constant.
The inverse reaction cross section $\sigma(\epsilon)$ is  calculated using the Wong's expression \cite{wong};
\begin{equation}
\sigma(\epsilon) = \frac{\hbar \omega R_{0}^{2}}{2\epsilon}ln(1+exp[\frac{2\pi}{\hbar \omega}(\epsilon -V_{B})]) ,
\label{eq:inverse}
\end{equation}
where $\hbar \omega$ is the curvature of fusion barrier for
angular momentum $\ell$=0. The V$_{B}$ is the emission barrier height of the protons, and it is
calculated using the expression \cite{yanez2008};
\begin{equation}
 V_{B}=\frac{1.44 Z_{P}(Z_{S}-Z_{P})} {r_{0}[A_{P}^{1/3}+(A_{S}-A_{P})^{1/3}]+\delta},
\label{barrier}
\end{equation}

where A$_{P}$,  Z$_{P}$  and A$_{S}$,  Z$_{S}$ are the mass and charge of the proton and emitting source, respectively. The value of $r_{0}$ is taken to be 1.45 fm \cite{ykg11}. $\delta$ is a factor which takes into account for the reduction in emission barrier due to deformation effects  and  it is taken to be 2.0 \cite{yanez2008} 
for compound nucleus  and 0.4 for fission fragments \cite{ikezoe1,alexender_vaz1982}. Thus, the emission barrier heights calculated for compound nucleus ($V^\mathrm{pre}_{B}$) and  FF ($V^\mathrm{post}_{B}$) are 11.1 and 7.6 MeV, respectively. The temperatures $T_\mathrm{pre}$ and $T_\mathrm{post}$ have been calculated using $T=\sqrt{E^{*}/a}$, where $E^{*}$ is the intrinsic excitation energy of the source and $a$ is the level-density parameter taken as A/11 for compound nucleus and A/7 for fission fragments \cite{ykg11, KR2006}. After correcting  the energy losses in the target, the intrinsic excitation energy of the source $E_\mathrm{CN}$ is 52.25 MeV for the present reaction. $T_\mathrm{pre}$ is scaled down  by a factor of 11/12 to account for multi-step evaporation \cite{KR2006,holub, couteur}. Thus, the temperatures $T_\mathrm{pre}$ and $T_\mathrm{post}$ values are calculated to be 1.40 and 1.46 MeV, respectively. It is worth to note here that the values of $T_\mathrm{pre}$ and $T_\mathrm{post}$, and the parameter, $\delta$, responsible for lowering of emission barrier due to deformation, are the same as used for $\alpha$-particle emission in the present heavy-ion induced fission reaction \cite{ykg22}.

The proton multiplicity spectra in the rest frames of three sources, namely, the compound nucleus (pre-scission), and both the FFs (post-scission)  were converted to laboratory frame using the appropriate Jacobians and finally summed up to fit the measured spectra. In the moving source fit, the parameters $V^\mathrm{pre}_{B}$, $V^\mathrm{post}_{B}$, $T_\mathrm{pre}$ and $T_\mathrm{post}$ are kept fixed, whereas the pre- and post-scission multiplicity are varied. The Total Kinetic Energy (TKE) release in the fission process is determined using Viola's systematics \cite{viola_1985} to be 182.5 MeV. The fitted spectra for the individual source and after summing are shown in the Fig. \ref{DataFit} for the $^{16}$O  + $^{232}$Th reaction at 10 typical angular settings. The best fitted values of the multiplicities are observed to be $\pi_\mathrm{pre}$ = (1.5 $\pm$ 0.3)$\times$10$^{-4}$, and  $\pi_\mathrm{post}$ = (6.8 $\pm$ 1.4)$\times$10$^{-5}$ corresponding to a minimum $\chi^{2}$/(degree of freedom) value of 6.4.


One can notice from  the Fig. \ref{DataFit} that the spectral fitting at relative angles around the perpendicular to the fission axis ($\theta_\mathrm{pf1} \approx 90^{\circ}$) as well as along the fission axis ($\theta_\mathrm{pf1} \approx 0^{\circ}$ \& $180^{\circ}$) is relatively poor; the fitted yield is substantial lower than the experimental data. The primary indication of extra yield at relative angles around parallel and perpendicular to the fission axis suggests the presence of near scission component, akin to polar- and equatorial-emission of protons in the low energy fission. A reanalysis of the data (MSDA) is carried out with inclusion of two near-scission sources (PE and EE) along with the pre- and post-scission sources. Similar to $\alpha$ particles, the energy and angular distributions for the equatorial protons are assumed to be Gaussian in the rest frame as given by expression \cite{ykg11, ykg12, ykg18};

\begin{equation}
\hspace{-0.8cm}
n(_{\epsilon, \theta})=N_{E} \pi_\mathrm{nseE}\exp\bigg[\frac{-(\epsilon-\epsilon_\mathrm{E}^\mathrm{p})^{2}} {2\sigma_\mathrm{\epsilon E}^{2}}\bigg] \exp\bigg[\frac{-(90^{\circ} - \theta_\mathrm{rel})^{2}} {2\sigma_\mathrm{\theta E}^{2}}\bigg]
\label{nseE}
\end{equation}
where, the $N_{E}$, $\pi_\mathrm{nseE}$, $\theta_\mathrm{rel}$, $\epsilon_\mathrm{E}^\mathrm{p}$, $\sigma_\mathrm{\theta E}$, and $\sigma_\mathrm{\epsilon E}$ are the normalization constant, proton multiplicity, relative angle of protons with respect to the scission axis, peak (or mean) energy, standard deviations of the angular, and energy distributions, respectively, in the rest frame for near-scission equatorial emission (EE). The polar emission in heavy-ion induced fission has been modeled in the MSDA for the first time. It is assumed that NSE protons originate around the neck region and attractive nuclear forces due to nascent fission fragment make a large fraction to be emitted along the scission axis. Thus, the energy and angular distributions for the polar protons can also be assumed to be Gaussian in the rest frame. It is given as; 
\begin{equation}
\hspace{-0.8cm}
n(_{\epsilon, \theta})=N_{P} \pi_\mathrm{nseP}\exp\bigg[\frac{-(\epsilon-\epsilon_\mathrm{P}^\mathrm{p})^{2}} {2\sigma_\mathrm{\epsilon P}^{2}}\bigg] \exp\bigg[\frac{-(\theta_{p} - \theta_\mathrm{rel})^{2}} {2\sigma_\mathrm{\theta P}^{2}}\bigg]
\label{nseP}
\end{equation}

where, $N_{P}$, $\pi_\mathrm{nseP}$, $\epsilon_\mathrm{P}^\mathrm{p}$, $\sigma_\mathrm{\theta P}$, and $\sigma_\mathrm{\epsilon P}$, are the normalization constant, proton multiplicity, peak (or mean) energy, standard deviations of the angular, and energy distributions, respectively, in the rest frame for near-scission polar emission. Here,  $\theta_\mathrm{p}$=0$^{\circ}$
if $\theta_\mathrm{rel}<=90^{\circ}$ and $\theta_\mathrm{p}$=180$^{\circ}$ if $\theta_\mathrm{rel}>90^{\circ}$.

During the reanalysis of the proton multiplicity spectra in the framework of MSDA, all four parameters related to each NSE component, PE \& EE, and multiplicities corresponding to pre- and post-scission sources were varied, and rest other parameters were kept fixed. The best fitted spectra for the individual source and after summing are shown in Fig. \ref{WithNSE_DataFit} for typical 10 angular settings. The best fitted values of the parameters are found to be $\pi_\mathrm{pre}$ = (1.5 $\pm$ 0.3)$\times$10$^{-4}$, $\pi_\mathrm{post}$ = (4.8 $\pm$ 1.1)$\times$10$^{-5}$, $\pi_\mathrm{nseE}$=(5.6 $\pm$ 0.8)$\times$10$^{-5}$, $\epsilon^{p}_{E}$ = 5.7 $\pm$ 0.3 MeV, $\sigma_{\epsilon E}$ = 2.1 $\pm$ 0.3 MeV, and $\sigma_{\theta E}$ = 8.2$^{\circ}$ $\pm$ $3.2^{\circ}$, $\pi_\mathrm{nseP}$=(4.6 $\pm$ 0.9)$\times$10$^{-5}$, $\epsilon^{p}_{P}$ = 4.5 $\pm$ 0.3 MeV, $\sigma_{\epsilon P}$ = 1.5 $\pm$ 0.3 MeV, and $\sigma_{\theta P}$ = 18.8$^{\circ}$ $\pm$ $3.1^{\circ}$ corresponding to a minimum $\chi^{2}$/(degree of freedom) value of 3.6. It is observed that with including both the components of NSE, the spectral fitting at angles around parallel and perpendicular to the scission axis improves significantly, whereas it remains almost the same at other angles. Also, $\pi_\mathrm{pre}$ and  $\pi_\mathrm{post}$ remain almost unaltered (within the experimental uncertainties) while the $\chi^{2}$/(degree of freedom) improves from 6.4 to 3.6 by including the PE and EE components of NSE. These observations reinforce the presence of near-scission proton emission in the present heavy-ion induced fission reaction. It is also seen that if the polar component is excluded, the spectral fitting around the relative angles parallel to the fission axis gets deteriorated along with worsening of $\chi^{2}$/(degree of freedom) from 3.6 to 4.2. Similarly, if the equatorial component is excluded, the spectral fitting around the relative angles perpendicular to the fission axis gets deteriorated along with worsening of $\chi^{2}$/(degree of freedom) from 3.6 to 6.1. It shows that polar and equatorial components are independent to each other.


\begin{figure}[t]
\centering\includegraphics [trim= 0.15mm 0.5mm 0.1mm 0.1mm,
angle=360, clip, height=0.27\textheight]{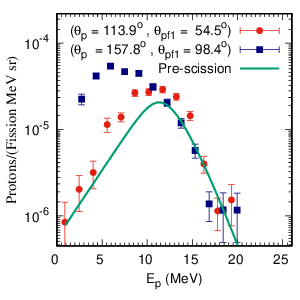} 
\caption{Proton multiplicity spectra in the center-of-mass frame of the compound nucleus at $\theta_{p}$=113.9$^{\circ}$, $\theta_\mathrm{pf1}$=54.5$^{\circ}$ (solid circles) and 
$\theta_{p}$=157.8$^{\circ}$, $\theta_\mathrm{pf1}$=98.4$^{\circ}$ (solid squares).  
The solid line represents the contribution from pre-scission emission in the center-of-mass frame.}
\label{Overlp}
\end{figure}

A comparison of proton multiplicity spectra, converted to center-of-mass frame of the compound nucleus, for two widely different laboratory angles; $\theta_{p}$=113.9$^{\circ}$, $\theta_\mathrm{pf1}$=54.5$^{\circ}$ (solid circles) and  $\theta_{p}$=157.8$^{\circ}$, $\theta_\mathrm{pf1}$=98.4$^{\circ}$ (solid squares) is shown in the Fig. \ref{Overlp}. Overlapping of the spectra (within the quoted uncertainties) beyond 12 MeV proton energy at above mentioned laboratory angles ensure that the compound nuclear evaporation dominates at these angular settings. The solid line in the Fig. \ref{Overlp} represents the contribution from pre-scission emission in the center-of-mass frame using the pre-scission multiplicity ($\pi_\mathrm{pre}$), determined using MSDA. The consistency of the above spectra with pre-scission component establishes the reliability of the energy calibration and the MSDA.


The parameters obtained from the MSDA for proton as well as $\alpha$-particle emission (reported in Ref. \cite{ykg22}) are compared in the Table \ref{table1}. It is interesting to note that both equatorial and polar components are observed to have similar intensities within the quoted uncertainties, however, the former is focused in a narrower cone ($\sigma_{\theta E}$ = 8.2$^{\circ}$) than the latter ($\sigma_{\theta P}$ = 18.8$^{\circ}$). It is known since a long time that equatorial emission is focused due to the Coulomb forces of the nascent fission fragments; the symmetric fission at higher excitation energy makes them well focused perpendicular to scission axis \cite{ykg11, KR2006}. On the other hand, it is the first time observation of polar proton emission in a heavy-ion induced fission. Present observations are qualitatively in agreement with theoretical prediction by Wada $et$ $al.$ \cite{Wada2015} who have reported angular distribution of scission particles within a simplistic approach. In Ref. \cite{Wada2015}, the angular distribution is obtained by integrating the time-dependent Schr\"{o}odinger equation using optical model for potential scattering (Coulomb and nuclear). It is shown that the angular distribution peaks at a relative angle close to the fission axis and  Coulomb field of both the fragments enhances the yield at $\theta_\mathrm{rel}$=90$^{\circ}$. The calculated angular widths of both the components are of similar magnitude ($\sim$15$^{\circ}$), whereas experimentally obtained, $\sigma_{\theta P}$ is observed close to twice of $\sigma_{\theta E}$. Calculations based on  more realistic model \cite{Carjan2013}, as has been applied so far only for scission neutrons \cite{Carjan2015, Carjan2019}, may reproduce the experimental angular widths.

\begin{figure}[t]
\centering\includegraphics [trim= 0.15mm 0.5mm 0.1mm 0.1mm,
angle=360, clip, height=0.35\textheight]{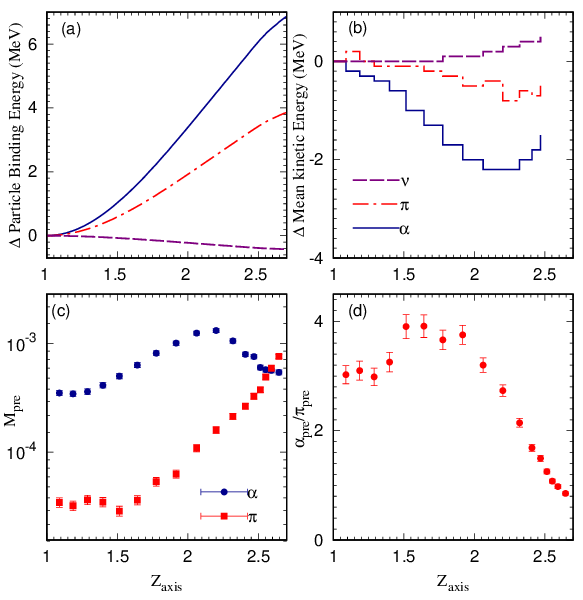} 
\caption{Change in binding energies (panel (a)) and mean kinetic energies (panel (b)) of neutrons, protons, and $\alpha$ particles with respect to spherical shape of the compound nucleus populated in the present reaction. In panel (c), the pre-scission multiplicities, of $\alpha_\mathrm{pre}$ and $\pi _\mathrm{pre}$ are shown as function of deformation while in panel (d) the ratio of $\alpha_\mathrm{pre}$ to $\pi _\mathrm{pre}$ is depicted as a function of deformation.} 
\label{KE_BE}
\end{figure}

It is seen from Refs. \cite{ykg11, ykg22, ykg12, ykg18} that $\alpha$-particle multiplicity spectra are well fitted without including the polar component of near-scission emission in heavy-ion induced fission. In fact, aforementioned calculations \cite{Carjan24} also indicate  absence of polar component for  $\alpha$ particles due to large absorptive potential (imaginary part of the optical potential). In the low energy fission, a small fraction for polar emission of  $\alpha$ particles has been reported \cite{PIASECKI1970}, however, their origin might be different.

The $\pi _\mathrm{pre}$  is observed to be close to one order lesser than the $\alpha_\mathrm{pre}$. Fractions of the NSE component with respect to the total pre-scission component (pre+NSE) are  provided in the Table \ref{table1}. This fraction for $\alpha$ particles is 11.4\%, consistent with the global systematics developed earlier \cite{ykg11}. Whereas, the same fraction for protons including both PE and EE is 40.7\%, showing a large fraction of pre-scission protons is emitted near to the scission stage. It is interesting to highlight here that total pre-scission multiplicity for protons is one order magnitude smaller than the $\alpha$ particles, however, their near scission fraction is around four times larger than the corresponding fraction for $\alpha$ particles.

 In the low-energy fission, the ratio of the ternary emission (equatorial) yield  of $\alpha$ particles to protons ($\alpha_\mathrm{nseE}$/$\pi _\mathrm{nseE}$)  is observed to be close to 100 \cite{nadkarni, halpern}. Had the emission mechanism at the scission stage be the same for low-energy fission and heavy-ion induced fission, the above ratio should have been observed close to 100 in the present reaction also. However, in the present reaction the same ratio is observed to be close to 5.5 (see Table \ref{table1}). This significant reduction in the near-scission (EE) yield  of $\alpha$ particles with respect to protons indicate about change of particle emission mechanism at the scission stage while going from low-energy to heavy-ion induced fission.

The yield ratio of pre-scission  $\alpha$ particles to protons is close to 16.2 which reduces to 3.0 while reaching to the scission stage as shown in the Table \ref{table1}. It is intriguing to understand that ratio of $\alpha$-particle to proton multiplicities decreases as we go from lower deformation (pre-scission) to very high deformation (near-scission). In addition to large deformation at the scission stage during the usual binary fission, the inverse trajectory calculations \cite{Pawan2024} show that the scission configuration is further stretched during the ternary fission leading to extraordinarily large deformation.

\begin{table}
 \centering
 \caption{\label{table1} Parameters extracted from MSDA for proton and $\alpha$-particle emission in $^{16}$O  + $^{232}$Th reaction at a beam energy of 96 MeV. The fractional NSE component is calculated as  $\pi_\mathrm{nse}$/($\pi_\mathrm{pre}$+$\pi_\mathrm{nse}$) for protons and  $\alpha_\mathrm{nse}$/($\alpha_\mathrm{pre}$+$\alpha_\mathrm{nse}$) for $\alpha$ particles.}
 \begin{tabular}{l c c rrrrrrr}

 \hline \hline 
 Parameter & Proton & $\alpha$-particle & ($\alpha$/p) \\ [1ex]
 \hline 

 M\footnotemark[0]$_\mathrm{pre}$ & (1.5$\pm$0.3)$\times$10$^{-4}$ & (2.4$\pm$0.2)$\times$10$^{-3}$  & 16.2$\pm$3.3 \\
 
 $T_\mathrm{pre}$ & 1.40 & 1.40 &   \\
 
 $V^\mathrm{pre}_{B}$ & 11.1 & 20.6 &    \\
 
 &  &  &    \\
 
 M\footnotemark[0]$_\mathrm{post}$ &  (4.8$\pm$1.1)$\times$10$^{-5}$ & (5.7$\pm$0.6)$\times$10$^{-5}$  & 1.2$\pm$0.3 \\
 
 $T_\mathrm{post}$ & 1.46 & 1.46 &   \\
 
 $V^\mathrm{post}_{B}$ & 7.6 & 13.7 &    \\
 
 &  &  &    \\
 
 M\footnotemark[0]$_\mathrm{nseE}$ & (5.6$\pm$0.8)$\times$10$^{-5}$ & (3.1$\pm$0.3)$\times$10$^{-4}$ &  5.5$\pm$1.0 \\
 
 $\epsilon^{p}_{E}$ & 5.7$\pm$0.3 & 15.4$\pm$0.4 &     \\
 
 $\sigma_{\epsilon E}$ & 2.1$\pm$0.3 & 3.6$\pm$0.3 &    \\
 
 $\sigma_{\theta E}$ & 8.2$\pm$3.2 & 10.7$\pm$1 &     \\
 
 &  &  &    \\
 
M\footnotemark[0]$_\mathrm{nseP}$ & (4.6$\pm$0.9)$\times$10$^{-5}$ &  &   \\
 
 $\epsilon^{p}_{P}$ & 4.5$\pm$0.4 &  &     \\
 
 $\sigma_{\epsilon P}$ & 1.5$\pm$0.3 &   &    \\
 
 $\sigma_{\theta P}$ & 18.8$\pm$3.1 &   &     \\ 
 
 &  &  &    \\
 
M\footnotemark[1]$_\mathrm{nseT}$ & (1.0$\pm$0.1)$\times$10$^{-4}$ & (3.1$\pm$0.3)$\times$10$^{-4}$  &  3.0$\pm$0.4\\

M\footnotemark[0]$_\mathrm{pre}$+M\footnotemark[0]$_\mathrm{nseE}$ & (2.0$\pm$0.3)$\times$10$^{-4}$ & (2.7$\pm$0.2)$\times$10$^{-3}$  &  13.3$\pm$2.1\\

M\footnotemark[0]$_\mathrm{pre}$+M\footnotemark[1]$_\mathrm{nseT}$ & (2.5$\pm$0.3)$\times$10$^{-4}$ & (2.7$\pm$0.2)$\times$10$^{-3}$  &  10.9$\pm$1.5\\ 

 Frac. of M\footnotemark[0]$_\mathrm{nseE}$  & (22.4$\pm$4.3)\% & (11.4$\pm$1.4)\% &   \\
 Frac. of M\footnotemark[0]$_\mathrm{nseP}$  & (18.3$\pm$4.4)\% &   &   \\
 Frac. of M\footnotemark[1]$_\mathrm{nseT}$  & (40.7$\pm$7.1)\% & (11.4$\pm$1.4)\% &   \\
 \hline \hline
 
 \end{tabular}
 \footnote[0]~ M refers to multiplicity \footnote[1]~Total NSE (EE + PE)
 \end{table}

Statistical model calculations were performed using the code JOANNE2 which incorporates the deformation dependent particle binding energies and transmission coefficients 
\cite{LestonePRL1993}. The changes in mean kinetic energies and binding energies of neutrons, protons, and $\alpha$ particles with respect to spherical shape of the compound nucleus populated in the present reaction are shown in the 
Fig. \ref{KE_BE} (a) and \ref{KE_BE} (b), respectively. One can note from Fig. \ref{KE_BE} (a) that the increase in binding energy of charged particles due to deformation is much more than that for neutrons. The lowering of emission barrier due to deformation for charged particles is of lesser magnitude than the increase in the binding energy (see Figs. \ref{KE_BE} (a) \& (b)). These two oppositely varying effects do not allow $\alpha_\mathrm{pre}$ to vary as much as the $\pi_\mathrm{pre}$ with increasing deformation  (see Fig.  \ref{KE_BE} (c)). The  $\pi_\mathrm{pre}$ keeps on increasing with increasing deformation; for the same excitation energy the emission probability for protons at near-scission  is more than the prior to scission. The ratio of $\alpha_\mathrm{pre}$ to $\pi _\mathrm{pre}$ is observed to be of decreasing nature with increasing deformation as shown in the Fig. \ref{KE_BE} (d). Thus, the decrease in experimentally observed multiplicity ratio ($\alpha$-particle to proton) from pre- to near-scission point can be attributed purely to the deformation effects while considering both as the statistical emissions. The increased fraction of near-scission protons (40.7\% including both PE and EE) in comparison to the same fraction for $\alpha$ particles (11.4\%) is qualitatively in agreement with above calculations. It implies that similar to ternary $\alpha$-particle emission in heavy-ion induced fission \cite{ykg11}, the near-scission protons are also statistically emitted. Whereas, in the low energy fission, the ternary particle emission is a dynamical process \cite{halpern}. Thus, the proton emission results presented in this Letter validate the conjecture made earlier from a systematic study of  $\alpha$-particle emission that neck rupture process makes a changeover from lower (for dynamical process) to higher viscosity (for statistical process) with increasing excitation energies. It is of further interest to determine the temperature at which such a transition occurs.

In summary, we have measured light charged particle energy spectra in coincidence with fission fragments in  $^{16}$O  + $^{232}$Th reaction at a beam energy of 96 MeV. The results from 
$\alpha$-particle emission have been reported earlier \cite{ykg22}. In the present work, proton multiplicity spectra are analyzed within the framework of a Moving Source Disentangling Analysis (MSDA) to determine contributions from different emission stages.  The MSDA conclusively shows ``near-scission proton emission" as an essential ingredient in the proton multiplicity spectra. These near-scission proton emissions are observed to be focused parallel (``polar") and perpendicular (``equatorial") to the fission axis with similar intensities. It is the first time that ``near-scission proton emission" has been disentangled from other emission stages in a heavy-ion induced fusion-fission process. In fact, inevitable presence of ``polar emission" for any type of particle in a heavy-ion reaction has been observed for the first time. It is also observed that pre-scission multiplicity for protons is one order of magnitude lesser than those of $\alpha$ particles. However, the near scission fraction of the proton yield is almost four times larger than the same fraction for $\alpha$ particles which is attributed to deformation effects of the fissioning nucleus. Present results open up a new avenue for simultaneous investigation of different particle emissions in heavy-ion induced fission reactions with varying degrees of freedom. By combining near-scission $\alpha$-particle and proton emissions in heavy-ion induced fission, it is reinforced that neck-rupture process during the scission stage makes a changeover from low to high viscosity in going from low (below say, 10 MeV) to high excitation energies (above say, 35 MeV). Specific investigations to such a transition point are of fundamental importance.

The authors would like to thank the operating staff of Pelletron accelerator
facility for the excellent operation of the machine.


\end{document}